\documentstyle[12pt]{article}

\begin{document}
\begin{center}
{\large Can internal shocks produce the variability in GRBs? }\\[0pt]
\vspace{0.4cm} Shiho Kobayashi\footnote{%
shiho@alf.fiz.huji.ac.il}, Tsvi Piran\footnote{%
tsvi@shemesh.fiz.huji.ac.il} and Re'em Sari\footnote{%
sari@shemesh.fiz.huji.ac.il}\\[0pt]
{\it {\small Racah Institute of Physics, Hebrew University, Jerusalem 91904,
Israel}} \vspace{0.8cm}
\end{center}

%%%%%%%%%%%%%%%%%%%%%%%%%%%%%%%%%%%%%%%%%%%%%%%%%%%%%%%%%%%
\begin{abstract}
We discuss the possibility that gamma-ray bursts result from internal shocks
in an ultra-relativistic matter. Using a simple model we calculate the
temporal structure and we estimate the efficiency of this process. In this
model the ultra-relativistic matter flow is represented by a succession of
shells with random values of the Lorentz factor. We calculate the shocks
that take place between those shells and we estimate the resulting emission.
Internal shocks can produce the highly variable temporal structure observed
in most of the bursts provided that the source emitting the relativistic
flow is highly variable. The observed peaks are in almost one to one
correlation to the activity of the emitting source. A large fraction of the
kinetic energy is converted to radiation. The most efficient case is when
an inner engine produces shells with comparable energy but very different
Lorentz factors. It also gives the most preferable temporal structure.
\end{abstract}
%%%%%%%%%%%%%%%%%%%%%%%%%%%%%%%%%%%%%%%%%%%%%%%%%%%%%%%%%%%%%%

\section{introduction}

The Burst and Transient Source Experiment (BATSE) on Compton-GRO has
revolutionized our ideas about the location of gamma-ray bursts(GRBs). The
isotropy of the events and the paucity of week bursts strongly support the
cosmological origin of GRBs (Meegan et al. 1992; Paczy\'nski 1992; Piran
1992; Nemiroff et al. 1994). The most fascinating mechanisms for producing
cosmological GRBs is decelerations of ultra relativistic matter having a
Lorentz factor $\gamma \ge 100$. This provides the only known solution to
the compactness problem (Piran 1996). The kinetic energy of the
ultra-relativistic matter is converted into internal energy by relativistic
shocks. These shocks can be due to the inter stellar medium (ISM),
``external shocks'' or inside the shell itself due to non-uniform velocity
``internal shocks''. Electrons are heated by the shocks and the internal
energy is then radiated via synchrotron emission and inverse Compton
scattering. Though the ultra relativistic matter flow was originally
considered in the dynamical context of fireball(Shemi \& Piran 1990;
Paczy\'nski 1990), one can imagine GRB models in which the fireball is
replaced by some unknown non-thermal acceleration mechanism but the
radiation still originate from slowing down of the ultra relativistic matter
(Piran 1996).

Most of bursts have a highly variable temporal profile with a time scale of
variability $\delta T$ significantly shorter than the overall duration $T$.
Typical value is $\delta T/T \sim 10^{-2}$. Clearly any GRB model must be
able to explain this complex temporal structure.

The naive external shocks scenario requires only one uniform shell. The
complex structure is due to surrounding matter (ISM in some models and star
light in others (Shaviv \& Dar 1995)). Two of us have shown that the
efficiency of this process is less than $1\%$ (Sari \& Piran 1997a). Using
the observed variability scale $\delta T$ one can estimate the typical size
of a single emitter. The number of emitters must be less than the number of
peaks. The total area of all emitters is therefore only a $\sim 1\%$ of the
shell's area. Even a modified version of this model, in which the shell is
composed of many smaller shells that collide one after the other with the
ISM is impossible. Most of the collisions would occur at a radius which
satisfies $R\ge \Delta \gamma ^{2}$ where $\Delta $ is the total width of
the shells\footnote{%
Distances, time, velocities $\beta$ and and the corresponding Lorentz
factors $\gamma$, are measured in the observer's rest frame. Thermodynamic
quantities are measured in the fluid rest frame.}. At this radius angular
spreading (Katz 1994; Fenimore et al. 1996) will cause each peak to be
spread on time scale of $R/\gamma ^{2}c=\Delta /c$ which is also the total
duration of the bursts. The observed bursts will therefor be smooth. The
only way out, in the context of the external shock model is if the shell is
replaced by a narrow and highly variable beam. However, the angular width of
this beam bust be than $\delta T/(T\gamma )\approx 10^{-4}$!

The obvious alternative is the internal shock scenario (Rees \&
M\'{e}sz\'{a}ros, 1994; Narayan, Paczy\'{n}ski \& Piran, 1992). These shocks
could occur within a variable relativistic wind produced by a highly
variable source. Internal shocks do not suffer from the angular spreading
problem. A fast shell that was injected after a slower one will eventually
catch up and collide with it. If both shells have Lorentz factor of order $%
\gamma $, the time until collisions will be of order of $\gamma ^{2}L/c$,
where $L$ is the initial separation between the shells. The collision will
therefor occur at radius $R=\gamma ^{2}L$ and the angular spreading time is
of order of $L/c$. Since the total duration is given by the total width of
the shells $\Delta /c\gg L/c$, the observed peak width is considerably
smaller than the total width and complex temporal structure is possible. In
view of the difficulties within the external shock scenario it is worthwhile
to consider the question whether internal shocks could actually produce the
temporal structure observed in GRBs and what would be the efficiency of this
process.

To examine the energy conversion in internal shock we construct, following
(Mochkovitch Maitia \& Marques, 1995) a simple model in which the
ultra-relativistic wind is represented by many shells having a random
distribution of intrinsic parameters. We calculate the energy conversion due
to shocks that form between the shells, considering at each time a binary
(two shells) encounter. We show that internal shocks can produce the
observed highly variable temporal profiles. We reproduce the result of
Mochkovitch, Maitia \& Marques (1995) who concluded that the efficiency of
this process is low $(<10\%)$ if the spread in Lorentz factors is small.
However, we show that higher efficiency could be achieved if the spread in
the Lorentz factors of the ultra-relativistic matter is larger.

We explore in section 2 the basic unit of our model, the interaction of two
shells. We explain in section 3 the algorithm for evolution of multiple
shells. In section 4 and 5, we discuss the temporal structure and the
efficiency, respectively. Finally we discuss the implication to observation
in section 6.

%%%%%%%%%%%%%%%%%%%%%%%%%%%%%%%%%%%%%%%%%%%%%%%%%%%%%%%%%%%%%%

\section{Two shell interaction}

Internal shocks arise in a relativistic wind with a non uniform Lorentz
factor and convert a portion of the kinetic energy to a radiation. We
represent the irregular wind by a succession of relativistic shells. A
collision of two shells is the elementary process in our model.

A rapid shell (denoted by the subscript $r$) catches up a slower one ($s$)
and the two merge to form a single one ($m$). The system behaves like an
inelastic collision between two masses $m_{r}$ and $m_{s}$. Using
conservation of energy and momentum we calculate the Lorentz factor of the
merged shell to be 
\begin{equation}
\gamma _{m}\simeq \sqrt{\frac{m_{r}\gamma _{r}+m_{s}\gamma _{s}}{%
m_{r}/\gamma _{r}+m_{s}/\gamma _{s}}},  \label{gammam}
\end{equation}
where $\gamma _{i}(\gg 1)$ and $m_{i}$ are the Lorentz factors and the
masses of these shells. The internal energy of the merged shell is the
difference of kinetic energy before and after the collision: 
\begin{equation}
E_{int}=m_{r}c^{2}(\gamma _{r}-\gamma _{m})+m_{s}c^{2}(\gamma _{s}-\gamma
_{m}).
\end{equation}
The efficiency of conversion of the kinetic energy into the internal energy
in a single collision is given by 
\begin{equation}
\epsilon =1-(m_{r}+m_{s})\gamma _{m}/(m_{r}\gamma _{r}+m_{s}\gamma _{s}).
\label{two-shell-efficiency}
\end{equation}

%If the mass is correlated to the Lorentz factor as we discuss 
%%in section 3,
%%$m_i \propto \gamma_i^{\eta}$, the energy is expressed by,
%%\begin{equation}
%%E_{int} \propto \gamma_r^{\eta+1}+\gamma_s^{\eta+1}
%%-(\gamma_r^{\eta}+\gamma_s^{\eta})
%%\sqrt{(\gamma_r^{\eta+1}+\gamma_s^{\eta+1})/
%%(\gamma_r^{\eta-1}+\gamma_s^{\eta-1})}.
%%\label{eq:etadep}
%%\end{equation}
%%where we used $\gamma_i \gg 1$.
%%%%%%%%%%%

The emitted radiation will be observed as a pulse with a width $\delta T$.
Three time scales, the cooling time, the hydrodynamic time, and the angular
spreading time determine $\delta T$ \footnote{%
Clearly, if the source is located at a high redshift $z$, all the observed
time scales are stretched by a factor $1+z$. However, as we are interested
mainly in the ration between the width of the peaks and the overall duration
and in the efficiency energy conversion this factor is unimportant.}. The
internal energy is radiated via synchrotron emission and inverse Compton
scattering. In most of cases, the cooling time scale of electrons is much
shorter than the hydrodynamic time scale (Sari et al. 1996; Sari \& Piran
1997b), so we can neglect the cooling time.

The hydrodynamic time scale is the time that the shock crosses the shell. In
fact there are two shocks as the interaction between the two shells takes
place in the form of two shocks: a forward shock and a reverse shock. Using
conservation of mass, energy and momentum at the shocks and equality of
pressures and velocities along the contact discontinuity, one can derive the
Lorentz factors of the forward and the reverse shocks $\gamma _{fs},\gamma
_{rs}$ (Sari \& Piran 1995): 
\begin{equation}
\gamma _{fs}\simeq \gamma _{m}\sqrt{\left( 1+\frac{2\gamma _{m}}{\gamma _{s}}%
\right) /\left( 2+\frac{\gamma _{m}}{\gamma _{s}}\right) },\ \ \ \gamma
_{rs}\simeq \gamma _{m}\sqrt{\left( 1+\frac{2\gamma _{m}}{\gamma _{r}}%
\right) /\left( 2+\frac{\gamma _{m}}{\gamma _{r}}\right) }.
\end{equation}
For simplicity, we estimate the emission time scale by the time that the
reverse shock crosses the rapid shell: 
\begin{equation}
\delta t_{e}=l_{r}/c(\beta _{r}-\beta _{rs}).
\end{equation}
where $l_{r}$ is a width of the rapid shell. The emitting region moves
towards the observer, the observed time scale is shorter by a factor of $%
1/2\gamma _{m}^{2}$. If both shells have Lorentz factor of order $\gamma $,
this observed time scale is of order $l_{r}/c$.

If the collision of the shells takes place at a large radius $R$, angular
spreading effects the width of the pulse. The separation of the shells is $L$%
, the effect of angular spreading on the pulse width is $\sim L/c$ as we
mentioned in section 1. If the separation $L$ is larger than the width of
the shells $l$, the pulse width $\delta T$ is determined by angular
spreading. The shape of the pulse become asymmetric with a fast rise and a
slower decline (Fig. 1) which GRBs typically show. The peak amplitude is
given by $E_{int}c/l$. If angular spreading is significant, the amplitude is
lower by a factor of $1-(1+l/L)^{-2}$. The observed luminosity, ${\cal L}$
is given by 
\begin{equation}
{\cal L}(t)=\cases{ 0, & ($t<0$), \cr h\left[ 1-1/(1+2\gamma_{m}^2ct/R)^2
\right], & ($0<t<\delta t_e/2\gamma_{m}^2$), \cr h\left[
{1/(1+(2\gamma_{m}^2 t-\delta t_e)c/R)^2} -{1/(1+2\gamma_{m}^2ct/R)^2}
\right],& ($t> \delta t_e/2\gamma_{m}^2$), \cr }
\end{equation}
where $h=E_{int}2\gamma _{m}^{2}/\delta t_{e}$.

We assume that the internal energy produced by the collision is radiated
isotropically in the shell's rest frame. We also assume that all the
internal energy is radiated rapidly. In this case we obtain at the end of
the collision a single cold shell whose Lorentz factor is $\gamma _{m}$,
given by Eq. \ref{gammam}. The shocks compress the initial shells and it
results in a thinner shell with a width $l_{m}$: 
\begin{equation}
l_{m}=l_{s}\frac{\beta _{fs}-\beta _{m}}{\beta _{fs}-\beta _{s}}+l_{r}\frac{%
\beta _{m}-\beta _{rs}}{\beta _{r}-\beta _{rs}}.
\end{equation}
The density is discontinuous through the contact discontinuity. For
simplicity we average the density: 
\begin{equation}
\rho _{m}=\frac{\rho _{r}l_{r}\gamma _{r}+\rho _{s}l_{s}\gamma _{s}}{%
l_{m}\gamma _{m}}.
\end{equation}
We describe the resulting shell as a homogeneous shell with a constant
density $\rho _{m}$ and a Lorentz factor $\gamma _{m}$.

While the shells propagate, their density decrease due to the spherical
geometry. However the density ratio between any two shells remain constant
and the process depends just on the density ratio and not on the absolute
value of the density. We expect therefor that this one dimensional model
captures the basic features of realistic events.

%%%%%%%%%%%%%%%%%%%%%%%%%%%%%%%%%%%%%%%%%%%%%%%%%%%%%%%%%%%%%%%%

\section{The Multiple shell model}

We represent the irregular wind by a succession of relativistic shells with
a random distribution of Lorentz factors and densities. A collision between
two shells produce shocks which convert some of the kinetic energy of the
shells to thermal energy. The thermal energy is then radiated via
synchrotron emission and inverse Compton scattering. In section 2, we
calculated the observed radiation produced by a collision between two
shells. For multiple shells, numerous collisions take place. Each collision
produces a pulse similar to that shown in Fig. 1. To construct the temporal
structure in a given model we calculate the time sequence of the two shell
collisions and we superimpose the resulting pulses from each collision.

Consider a wind consisting of $N$ shells. We assign an index $i$, ($i=1,N$)
to each shell according to the order of the emission from the inner engine.
Each shell is characterized by four variables: a Lorentz factor $\gamma _{i}$%
, a density $\rho _{i}$, (the shells are cold and therefore the pressure
satisfies $p_{i}\equiv 0$.) a width $l_{i}$ and the time, $\tilde{t}_{i}$,
when the shell was ejected from the source. We choose the origin of time
such that the last shell - the N-th shell - is emitted at $\tilde{t}_{N}=0$.
Clearly all other shells are emitted earlier so the $\tilde{t}_{i}$'s are
negative for $i<N$. At $t=0$ the shells are at the following ``initial''
positions (inner edge): $R_{i}=-\tilde{t}_{i}\beta _{i}c$. We denote the
distance at this stage between the shell $i$ and the shell $i+1$ by $L_{i}$: 
$L_{i}\equiv R_{i}-R_{i+1}-l_{i+1}$. The time $l_{i}/c$ corresponds to the
period that the ``inner engine'' has operated, while $L_{i}/c$ corresponds
to the period for which the ``inner engine'' was quiet. The size of the
inner engine should be, of course, smaller than $l$ to produce a shell with
a width $l$.

We turn now to evaluate the evolution of this multiple shell system. There
will be numerous collisions between different shells. We denote by the index 
$j$ the $j$-th collision which takes place at a time $t_{j}$ and position $%
R_{c}(t_{j})$. For simplicity we define a zeroth collision at the initial
time $t_{0}=0$ this enables us to put the first collision on par with the
rest. Given the velocities of the shells $\beta _{i}$ and their separation $%
L_{i}$ at $t_{j-1}$ we calculate for all pairs $(i,i+1)$ satisfying $\beta
_{i+1}>\beta _{i}$ the collision time: 
\begin{equation}
\delta t_{i,i+1}\equiv L_{i}/c(\beta _{i+1}-\beta _{i}).
\end{equation}
We then find the minimal collision time among all the $\delta t_{i}$'s: 
\begin{equation}
\delta t_{j}={\rm min}[\delta t_{i,i+1}].
\end{equation}
Let $s$ and $s+1$ be the index of the shells for which this minimum takes
place. The $j$-th collision is between the $s$ and $s+1$ shells and takes
place at the time 
\begin{equation}
t_{j}=t_{j-1}+\delta t_{j}
\end{equation}
and at the position 
\begin{equation}
R_{c}(t_{j})=R_{s+1}(t_{j})=R_{s+1}(t_{j-1})+c\beta _{s+1}\delta t_{j}.
\end{equation}
An observer at $R_{0}$ away from the central source will begin to detect
radiation from this collision at a time 
\begin{equation}
t_{obs,s}=(R_{0}-R_{c}(t_{j}))/c+t_{j}.
\end{equation}
Note that, for reasons that will be clear later we give this observed time
the index $s$ of the outer shell that participated in the collision. At the
end of the computation we set the minimum of $t_{obs,s}$ to be the origin of
the observer time as this is when the first radiation is observed. We assign
the pulse shape presented in section 2 beginning at this time to the overall
observed signal.

We rearrange the shells after each collision as follows. Each shell moves
from its earlier position $R_{i}(t_{j-1})$ to 
\begin{equation}
R_{i}(t_{j})\equiv R_{i}(t_{j-1})+c\beta _{i}\delta t_{j}
\end{equation}
For the first collision $t_{j-1}=0$ and we use the ``initial positions''
here. All the shells except the $s$ shell and the $s+1$ shell keep their
Lorentz factor, the density and the width: $\gamma _{i}$, $\rho _{i}$ and $%
l_{i}$. The $s$ and the $s+1$ shell merge to form a single shell which we
label $s+1$. We delete the $s$ shell and we will skip from now on this index
when it will be encountered. Using the results of section 2 we calculate now
the new parameters for the $s+1$ shell: $\gamma _{s+1}$, $\rho _{s+1}$ and $%
l_{s+1}$.

We return now to the calculation of the distances between shells: $L_{i}$
and the corresponding collision time $\delta t_{i}$. We find the next
collision and proceed until there are no more collisions, i.e., until all
the shells have merged to form a single shell or until the shells are
ordered with increasing values of the Lorentz factor.

In order to reduce the number of parameters describing the wind we adopt the
following simplifications. We assume a constant initial width $l_{i}=l$ and
constant initial separation $L_{i}=L$. In other words, we assume that the
``inner engine'' operates for a fixed period $l/c$ and then it is quiet for
a fixed period $L/c$.

We assume that the Lorentz factor of each shell is uniformly distributed
between $\gamma _{min}$ and $\gamma _{max}$. For the distribution of the
density we have two kinds of models. In the first kind we consider models in
which the density, the mass or the energy of each shell is a random variable
which is uniformly distributed between $1$ and some maximal value $X_{max}$.
In the second kind the density is correlated with the Lorentz factor as $%
\rho _{i}\propto \gamma _{i}^{\eta -1}$ so that the mass and kinetic energy
of each shell are proportional to $\gamma _{i}^{\eta }$ and $\gamma
_{i}^{\eta +1}$ respectively.. For $\eta =1$ the shells have equal density,
for $\eta =0$ the shells have equal mass and for $\eta =-1$ the shells have
equal energy. We will see later that the random density/mass/energy models
give almost same efficiency as the corresponding constant model, and the
dependence on $X_{max}$ is low. We therefore present most of the result for
the correlation models which are more simple.

Any time scale in this system is proportional to a sum of $L$ and $l$, the
amplitudes of observed peaks depends neither on the absolute value of $l$
nor that of $L$. If both of $L$ and $l$ are transformed by the same factor,
we will get similar temporal profile. We are interested in neither the
absolute value of the duration nor that of the peak, then we can use a
single parameter $L/l$ instead of the two parameters $L$ and $l$. The
efficiency of the energy conversion by the shocks is independent of $L/l$.
The total energy emitted is the sum of the energy emitted in the elementary
two shell collisions which are independent of $L/l$. The density $\rho $ is
the only quantity in this system with dimensions of mass. Therefore only
density ratio rather than density can determine the time scale, and the
overall normalization of the density $\rho $ is also unimportant.
Consequently we find that there are only five initial parameters: $N$, $L/l$%
, $\gamma _{max}$, $\gamma _{min}$ and either $\eta $ (if the density is
correlated with the Lorentz factor) or $X_{max}$ (for random distribution of 
$\rho $,$m$ or $E$).

\section{Temporal Structure}

The calculated temporal structure is a superposition of the pulses from the
elementary two shell collisions. Several typical temporal profiles are
presented in Figs. 2a-2f. Fig. 2a depicts the luminosity for a model with $%
N=100$ and $L/l=5$ as a function of the observed time for shells with a
constant energy ($\eta =-1$). During the evolution of the $N$ shells, almost 
$N$ collisions happen and almost $N$ pulses are produced. Though some of the
peaks are not observed because of their higher neighbors. The number of
peaks is given by $\sim N$. In Fig. 2b the luminosity is plotted with the
same parameters as Fig. 2a except that $N=20$. Fig 2a and 2b show that $N$
practically determines the number of peaks. The width of a given peak $%
\delta T$ is determined by the emission time scale $\sim l/c$ or by the
angular spreading $\sim L/c$. If the value of the separation $L$ is larger
than that of the width $l$, $\delta T/T$ is $\sim 1/N$ and independent of $%
L/l$. In Fig. 2c, the luminosity is plotted with the same parameters as Fig.
2b except that $L/l=1$. The profile is almost independent of $L/l$.

The different amplitude of the peaks originate mainly from the difference in 
$E_{int}$, since all peaks have width of the same order of magnitude $\sim
L/c$. For given random Lorentz factors, $E_{int}$ depends on the index $\eta 
$ for models with density correlated to the Lorentz factor. The random
Lorentz factors are uniformly distributed between $\gamma _{min}$ and $%
\gamma _{max}$. If $\gamma _{max}/\gamma _{min}\gg 1$, the ratio $\gamma
_{r}/\gamma _{s}$ is typically large. In such a case, we obtain a simple
formula, 
\begin{equation}
E_{int}\sim \cases{ (1-1/\sqrt{2})\gamma_r^2, & ($\eta=1$), \cr 2-\sqrt{2},
& ($\eta=-1$). \cr }
\end{equation}
For $\eta =-1$, $E_{int}$ takes almost the same value for most of the
collisions so most of the peaks have comparable amplitude (see Fig. 2a).
Thus, for $\eta =-1$ we practically observe the ``shell structure'' produced
by the inner source as almost all two shell collisions produce an observable
peak. If on the other hand $\eta =1$ (see Fig. 2d) a few collisions produce
significantly higher peaks than others and only those are observed. Thus the
number of the observed peaks is much less than $N$, the observed temporal
structure corresponds only weakly to the activity of the inner source.
Visual inspection suggests that the temporal structure produced by $\eta =-1$
resembles better the observed temporal structure. 

Fig. 2e and Fig. 2f present the luminosity profiles for model with random
energy and random density respectively.. Both use $X_{max}=1000$ and both
with spread of Lorentz factor $\gamma _{max}/\gamma _{min}=10$. Again the
random density is more spiky than the random energy case. However as could
be guessed intuitively, the random energy and the random density cases are
more spiky than the constant energy ($\eta =-1$, Fig. 2a) and constant
density ($\eta =1$, Fig. 2d) respectively.. 

The total width of the wind is $\Delta =Nl+(N-1)L$. Neighboring shells
collide on a time scale of $L\gamma ^{2}/c$, with fluctuations of the same
order of magnitude due to the fluctuations of $\gamma $. The matter is
moving toward the observer, the resulting observed time scale is $L/c$. On
the other hand, the difference in observed time due to the location of the
given shell within the wind is a order of $\Delta /c$. Since by definition $%
\Delta \gg L$ we observe pulses arising from collision between shells mostly
according to their positions inside the wind, i.e., according to the time
that those shells were emitted by the inner engine. In Fig. 3, we plot the
time that radiation is observed from a shell $t_{obs,s}$ versus the time
that the shell was ejected from the source $\tilde{t}_{s}$. We associated
the emission from a given collision with the slower shell. The clear
correlation between the observed time and the time that the source ejected
the specific shell shows that the observed temporal structure reflects the
activity of the source. This conclusion is valid even if there is a large
spread in $\gamma $.

\section{Efficiency}

We have shown that internal shocks could produce the highly variable
temporal structure. We ask now what is the efficiency of this process? The
relativistic shells collide with each other and merge into more massive
shells. The overall efficiency of conversion of kinetic energy in internal
shocks can be calculated from the initial and final kinetic energies as: 
\begin{equation}
\epsilon =1-\frac{\Sigma m_{i}^{(f)}\gamma _{i}^{(f)}}{\Sigma
m_{i}^{(i)}\gamma _{i}^{(i)}}  \label{efi}
\end{equation}
where the superscript $(f)$ and $(i)$ represent the initial and final
values, respectively.

This efficiency depends on the parameters of the model $N,\gamma _{min}$, $%
\gamma _{max}$ and $\eta $ (or $X_{max})$ and on the specific realization:
the set of random Lorentz factors that are assigned to each shell. For each
choice of the parameters of the model, we have evaluates the efficiency for
100 realizations. The mean efficiency and its standard deviation are listed
in Table 1.

The efficiency is only a few percents if the spread in the Lorentz factor is
relatively low (a factor of 2-3). This agrees with results of Mochkovitch,
Maitia and Marques (1995) who concluded that the efficiency of this process
is less than $10 \%$. However, higher efficiency could be reached if the
spread in $\gamma$ is larger. The spread required, for example, for $10\%$
is $\gamma_{max}/\gamma_{min}\sim 6$ for $N=100$ and $\eta=-1$.

The efficiency is independent of $L/l$, which is important only for the
temporal structure. For models with density correlated with the Lorentz
factor the relevant parameters that determine the efficiency are $N$, $\eta $%
, $\gamma _{min}$ and $\gamma _{max}$ (see Table 1a). For the random density
models, we evaluated the efficiency with $X_{max}=1$ to $1000$. The
efficiency is almost independent of $X_{max}$. It can be seen that the
efficiencies of the models with random $n$, $m$ and $E$ are very similar to
those with $\eta =1,0$ and $-1$, respectively (Table 1b).

As the number of shells, $N$, increases, the kinetic energy of relative
motion of the final mergers becomes negligible. The efficiency approaches an
asymptotic value. Further if we consider large values of $\gamma _{min}$,
the efficiency depends only on the ratio $\gamma _{max}/\gamma _{min}$
reaching an asymptotic value for large $\gamma _{max}/\gamma _{min}$ (Fig.
4). For models with correlation between the density and the Lorentz factor,
this asymptotic value depends only on the index $\eta $. In the limits $\eta
\to \pm \infty $, a single shell carries all the energy of the system and
the energy and momenta of the other shells are negligible. Therefore, the
massive shell is unaffected by collisions with the lighter ones. We expect a
low efficiency in this limit (Fig. 5).

In order to understand the dependence of the efficiency on $\eta $ we
consider a simplified model. Let all the shell collide and merge into a
single shell and only then emit the thermal energy as the radiation. Using
conservation of energy and momentum we can calculate the Lorentz factor and
the efficiency: 
\begin{equation}
\gamma =\sqrt{\Sigma \gamma _{i}^{\eta +1}/\Sigma \gamma _{i}^{\eta -1}},
\end{equation}
\begin{equation}
\epsilon =1-\Sigma \gamma _{i}^{\eta }/\sqrt{\Sigma \gamma _{i}^{\eta
-1}\Sigma \gamma _{i}^{\eta +1}}.
\end{equation}
Averaging over the random variables $\gamma _{i}$, and assuming a large
number of shells $N\to \infty $ we obtain: 
\begin{equation}
\langle \epsilon \rangle \sim 1-\frac{(\gamma _{max}/\gamma _{min})^{\eta
+1}-1}{\eta +1}\sqrt{\frac{\eta (\eta +2)}{((\gamma _{max}/\gamma
_{min})^{\eta }-1)((\gamma _{max}/\gamma _{min})^{\eta +2}-1)}}.
\label{analitic}
\end{equation}
This formula explains qualitatively the behavior of our numerical results. A
comparison between the simulations in which the radiations is emitted right
after each collision and the simplified single merge model is presented in
Fig. 5b.

An interesting point to notice is that for many shell collision (both in the
numerical simulation and the simple model Eq. \ref{analitic}) the most
efficient case is where the shells have the same energy while for the two
shell collision (Eq. \ref{two-shell-efficiency}) the highest efficiency is
for shells with equal mass rather than equal energy.

%%%%%%%%%%%%%%%%%%%%%%%%%%%%%%%%%%%%%%%%%%%%%%%%%%%%%%%%%%%%%%%%

\section{Conclusions}

Using a simple model, we have shown that internal shocks can produce the
highly variable profile observed in most GRBs. There is a strong correlation
between the time at which we observe a pulse and the emission time of the
corresponding shell from the inner engine. This correlation persists even
when there is a large spread in the Lorentz factor. Thus the observed
temporal structure reproduces the activity of the source.

We have shown that the number of peaks is almost the same as the number of
shells that the inner engine emitted. The separation between the peaks
corresponds to the duration for which the inner engine was quiet. The
variability of peaks height can tell as wether all shells have comparable
energy (low variability in observed peak heights) or not. A systematic
statistical study comparing the temporal structure produced by internal
shocks with observations will be carried out in a subsequent paper.

The efficiency of this process is low (less than $2\%$) if the initial
spread in $\gamma $ is only a factor of two. However the efficiency could be
much higher. The most efficient case is when the inner engine produces
shells with comparable energy but with very different Lorentz factors. In
this case ($\eta =-1$, and spread of Lorentz factor $\gamma _{max}/\gamma
_{min}>10^{3}$) the efficiency is as high as $40\%$. For a moderate spread
of Lorentz factor $\gamma _{max}/\gamma _{min}=10$, with $\eta =-1$, the
efficiency if $20\%$. 

%%%%%%%%%%%%%%%%%%%%%%%%%%%%%%%%%%%%%%%%%%%%%%%%%%%%%%%%%%%%%%%%

We thank J.I.Katz for many useful discussion. S.K. gratefully acknowledges
the support by the Golda Meir Postdoc fellowship. This work was supported in
part by a US-Israel BSF grant and by a NASA grant.

%%%%%%%%%%%%%%%%%%%%%%%%%%%%%%%%%%%%%%%%%%%%%%%%%%%%%%%%%%%%%%%%
\newpage \noindent Fenimore,E.E., Madras, C.D., \& Nayakshin, S., 1996, {\it %
Ap.J.}, {\bf 473}, 998.\newline
Katz, J.I. 1994, {\it Ap. J.}, {\bf 432}, L107.\newline
Meegan, C.A. et al. 1992, {\it Nature}, {\bf 355}, 143.\newline
Mochkovitch, R., Maitia, V., \& Marques, R. 1995, in Towards the Source of
Gamma-Ray Bursts, Proceeding of 29th ESLAB Symposium, eds. Bennett, K. \&
Winkler, C., 531.\newline
Nemiroff, R.J., Noriss, J.P., Kouveliotou, C., Fishman, G.J., Meegan, C.A.,
\& Paciesas, W.S. 1994, {\it Ap. J.}, {\bf 423}, 432.\newline
Narayan, R., Paczy\'{n}ski, B., \& Piran, T. 1992, {\it Ap. J.}, {\bf 395},
L83.\newline
Paczy\'{n}ski, B. 1990, {\it Ap. J.}, 363, 218.\newline
Paczy\'{n}ski, B. 1992, {\it Nature}, {\bf 355}, 521.\newline
Piran, T. 1992, {\it Ap. J.}, {\bf 389}, L45.\newline
Piran, T. 1996, in Some unsolved problems in Astrophysics, eds. Bahcall. J.
\& Ostriker J.P., (Princeton University Press).\newline
Rees, M. J., \& M\'{e}s\'{z}aros, P. 1994, {\it Ap. J.}, {\bf 430}, L93.%
\newline
Sari, R. \& Piran,T. 1995, {\it Ap.J.}, {\bf 455}, L143.\newline
Sari,R., Narayan,R. \& Piran, T. 1996, {\it Ap.J. }, {\bf 473}, 204.\newline
Sari, R., \& Piran, T. 1997a, {\it Ap.J.} in press. \newline
Sari, R., \& Piran, T. 1997b, {\it MNRAS} in press.\newline
Shemi, A., \& Piran, T. 1990, {\it Ap.J.}, 365, L55.\newline
Shaviv, N., \& Dar, A. 1995, {\it MNRAS}, {\bf 277}, 287.\newline

\newpage

\begin{center}
\tabcolsep=3pt \footnotesize
Table.1 a\\
\begin{tabular}{|r|c|c|c|c|} \hline
$N$ & $\eta $ &$\gamma_{min}$& $\gamma_{max}$ &efficiency [\%]\\ \hline
100   & 1   &  20 & 1000 & $10.9\pm1.4$\\ 
100   & 1   &  50 & 1000 & $10.0\pm1.3$\\ 
100   & 1   & 100 & 1000 & $ 8.5\pm1.1$\\ 
100   & 1   & 200 & 1000 & $ 6.1\pm0.7$\\ 
100   & 1   & 500 & 1000 & $ 1.7\pm0.2$\\ \hline
100   & 0   &  20 & 1000 & $19.2\pm2.9$\\ 
100   & 0   &  50 & 1000 & $16.0\pm2.3$\\ 
100   & 0   & 100 & 1000 & $12.4\pm1.6$\\ 
100   & 0   & 200 & 1000 & $ 7.7\pm0.9$\\ 
100   & 0   & 500 & 1000 & $ 1.8\pm0.2$\\ \hline
100    & -1  &  20 & 1000 & $25.0\pm3.5$\\ 
100   & -1  &  50 & 1000 & $19.5\pm2.4$\\ 
100   & -1  & 100 & 1000 & $14.1\pm1.6$\\ 
100   & -1  & 200 & 1000 & $ 8.2\pm0.9$\\ 
100   & -1  & 500 & 1000 & $ 1.8\pm0.2$\\ \hline
20    & 1   & 100 & 1000 & $ 7.6\pm2.5$\\ 
20    & -1  & 100 & 1000 & $11.0\pm3.3$\\ \hline
1000  & 1   & 100 & 1000 & $ 8.7\pm0.3$ \\
1000  & -1 & 100 & 1000 & $15.1\pm0.5$ \\ 
1000  & -1 & 10 & 10000 & $39.3\pm2.6$ \\ \hline
\end{tabular}

\vspace{0.5cm}
Table.1 b\\
\begin{tabular}{|r|c|c|c|c|c|} \hline
$N$ & $  $ &$\gamma_{min}$& $\gamma_{max}$&$X_{max}$&efficiency [\%]\\ \hline
100  & random $n$ & 100 & 1000 &1000& $8.2\pm1.2$ \\ 
100  & random $m$ & 100 & 1000 &1000& $12.2\pm1.6$ \\ 
100  & random $E$ & 100 & 1000 &1000& $14.2\pm1.6$ \\ \hline
\end{tabular}

\vspace{0.5cm}

Mean efficiency and standard deviation of the efficiency 
for 100 realizations of different models. 
\end{center}

\newpage 
\noindent {\bf Figure captions}

\vspace{0.5cm}

Fig. 1 A peak produced by a collision between two shell. The luminosity
plotted versus the observer's time. The solid line corresponds to $R=c\delta
t_e $ and the dotted line corresponds to $R=0$.

\vspace{0.5cm}

Fig. 2 The observed temporal structures. The luminosity versus the observer's
time, for different models\\[0pt]
a: $\gamma_{min}=100$, $\gamma_{max}=1000$, $N=100$, $\eta=-1$ and $L/l=5$ 
\\[0pt]
b: $\gamma_{min}=100$, $\gamma_{max}=1000$, $N=20$ , $\eta=-1$ and $L/l=5$ 
\\[0pt]
c: $\gamma_{min}=100$, $\gamma_{max}=1000$, $N=20$ , $\eta=-1$ and $L/l=1$ 
\\[0pt]
d: $\gamma_{min}=100$, $\gamma_{max}=1000$, $N=100$, $\eta=1$ and $L/l=5$ 
\\[0pt]
e: $\gamma_{min}=100$, $\gamma_{max}=1000$, $N=100$,\\[0pt]
\ \ \ \ random energy with $E_{max}=1000$ and $L/l=5$ \\[0pt]
f: $\gamma_{min}=100$, $\gamma_{max}=1000$, $N=100$,\\[0pt]
\ \ \ \ random density with $\rho_{max}=1000$ and $L/l=5$

\vspace{0.5cm}

Fig. 3 The time of ejection of a shell by the inner engine, $\tilde t_j$
versus the observed time of the photon produced in that shell, $t_{obs,j}$,
for $N=100$, $\gamma_{min}=10$, $\gamma_{max}=1000$, $\eta=-1$ and $L/l=5$.
The initial position 0 and 100 correspond to the inner and outer edge of the
wind.

\vspace{0.5cm}

Fig. 4 The efficiency versus $\gamma_{max}/\gamma_{min}$ with $1\sigma$
error bars of 100 random simulations, for $\gamma_{min}=10$, $N=500$ and $%
\eta=-1$.

\vspace{0.5cm}

Fig. 5 The efficiency plotted versus the index $\eta$, for $\gamma_{min}=100$%
, $N=100$. The solid line corresponds to $\gamma_{max}/\gamma_{min}=1000$,
the dotted line corresponds to $\gamma_{max}/\gamma_{min} =100$ and the
dashed line corresponds to $\gamma_{max}/\gamma_{min}=10$. a: Numerical
simulation, b: Approximation by analytic formula (\ref{analitic})

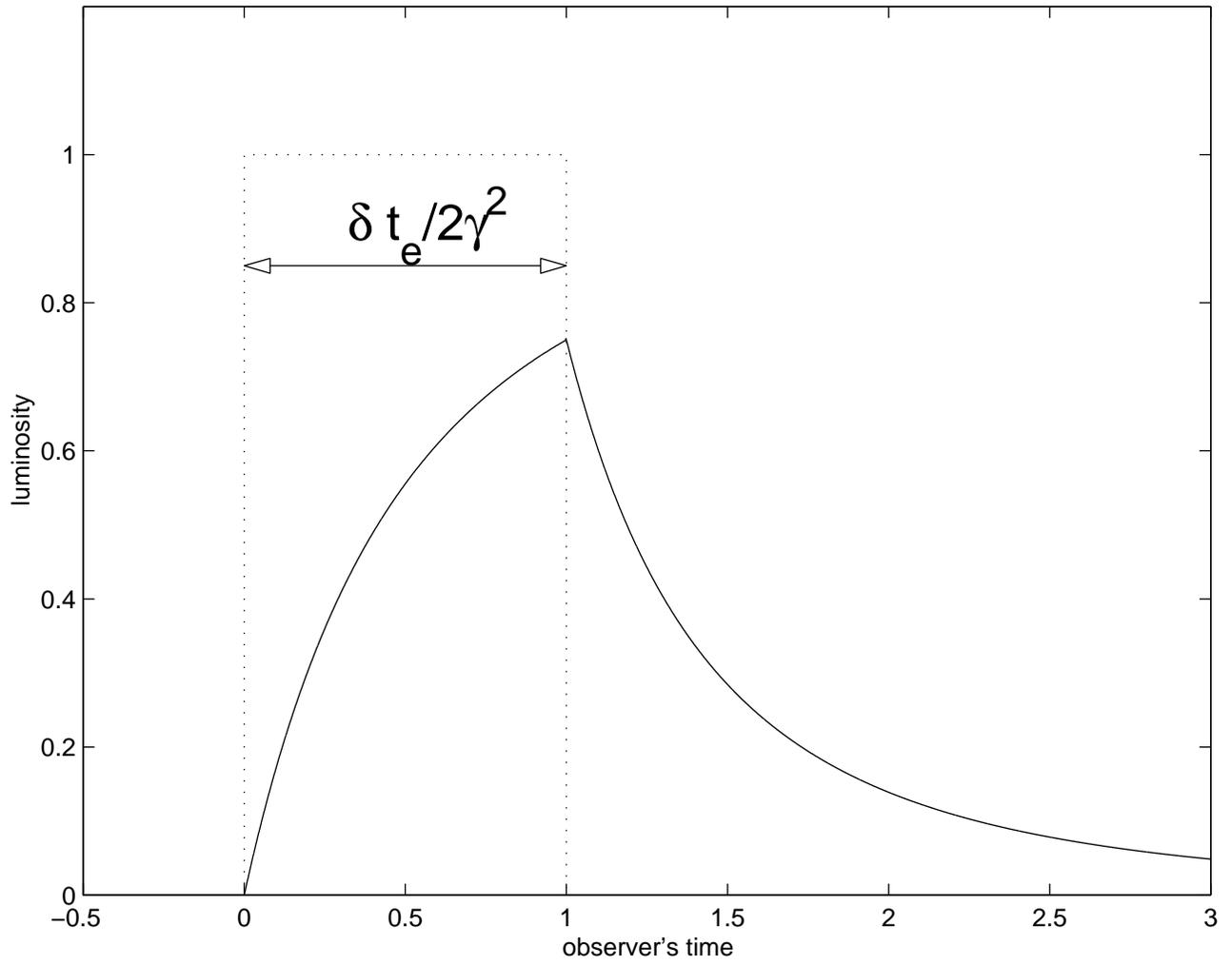
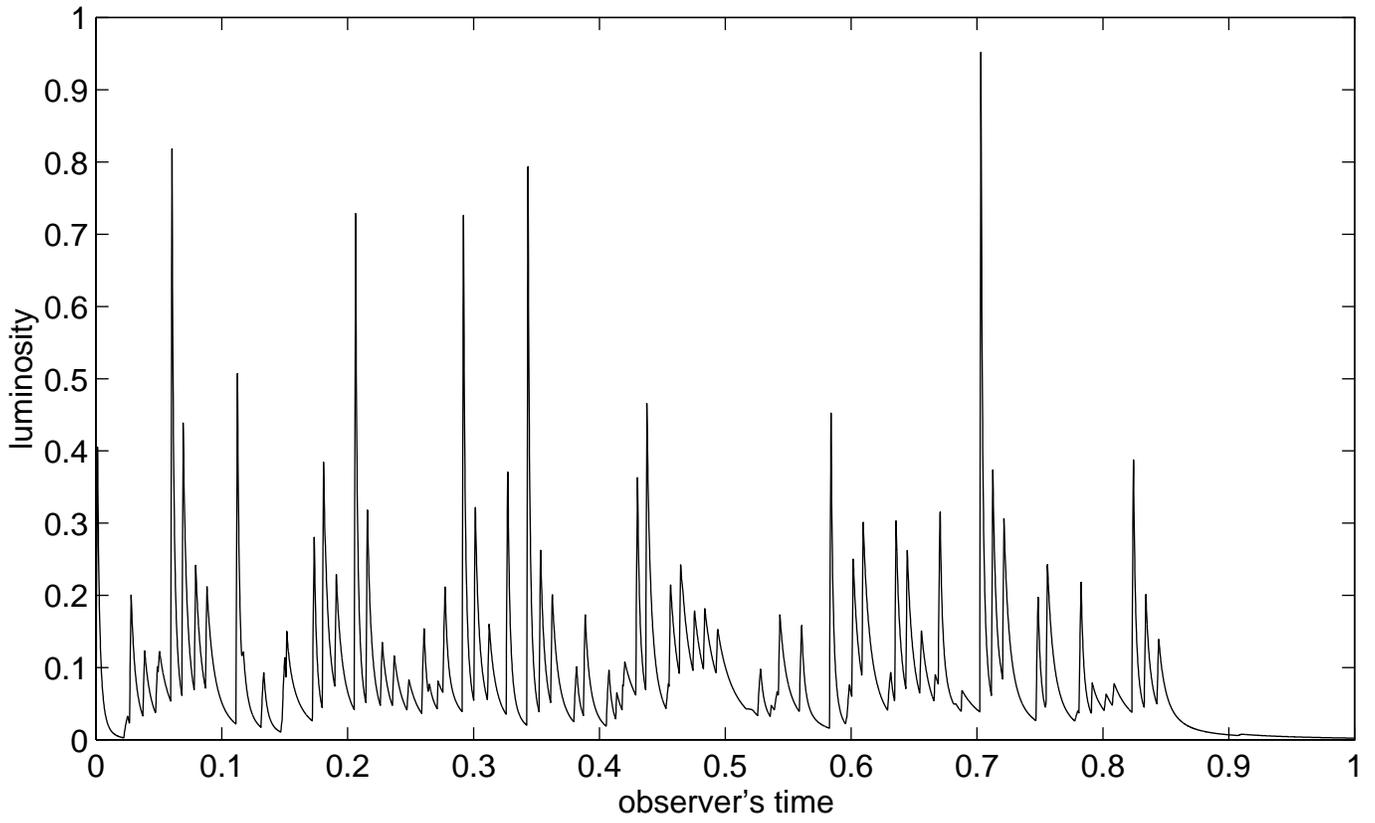
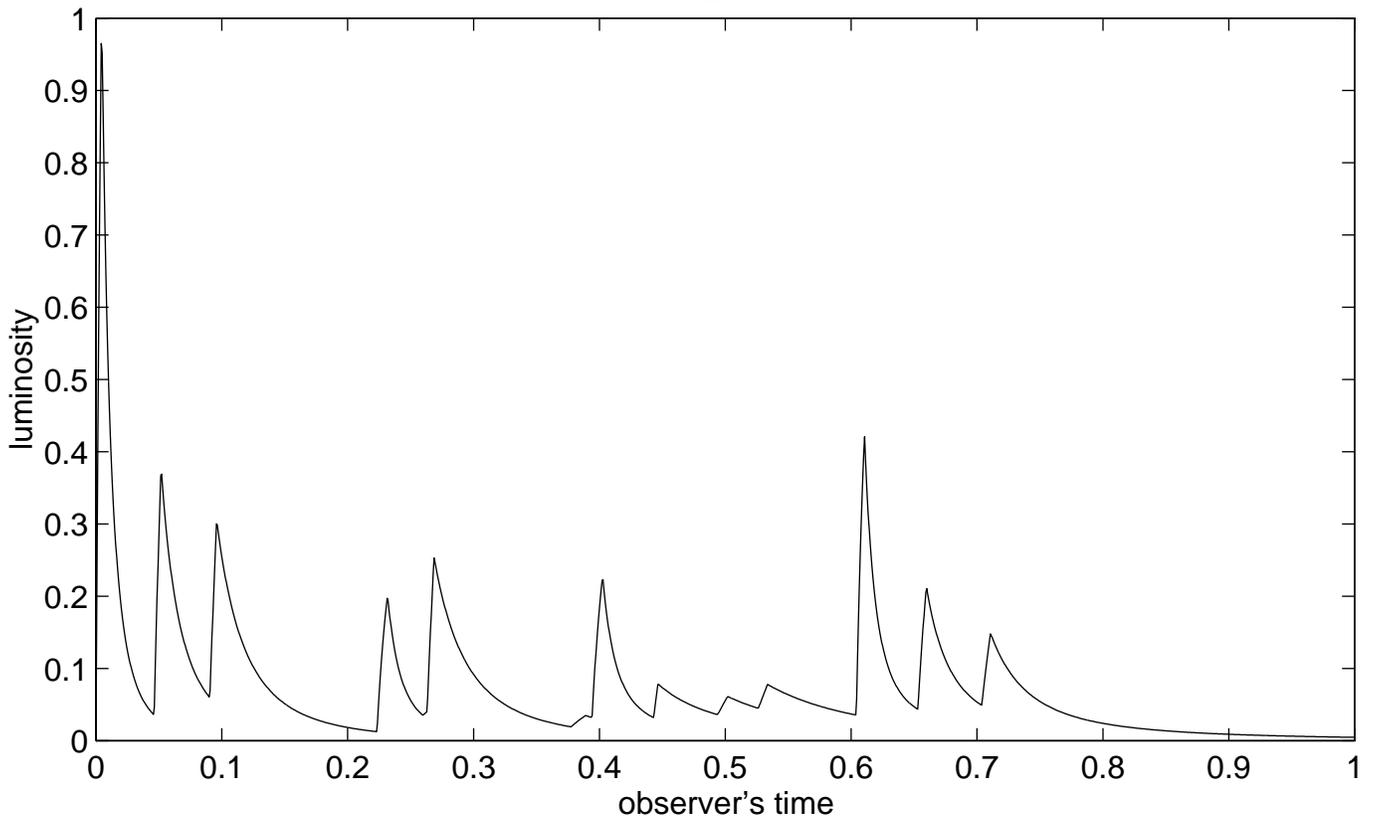
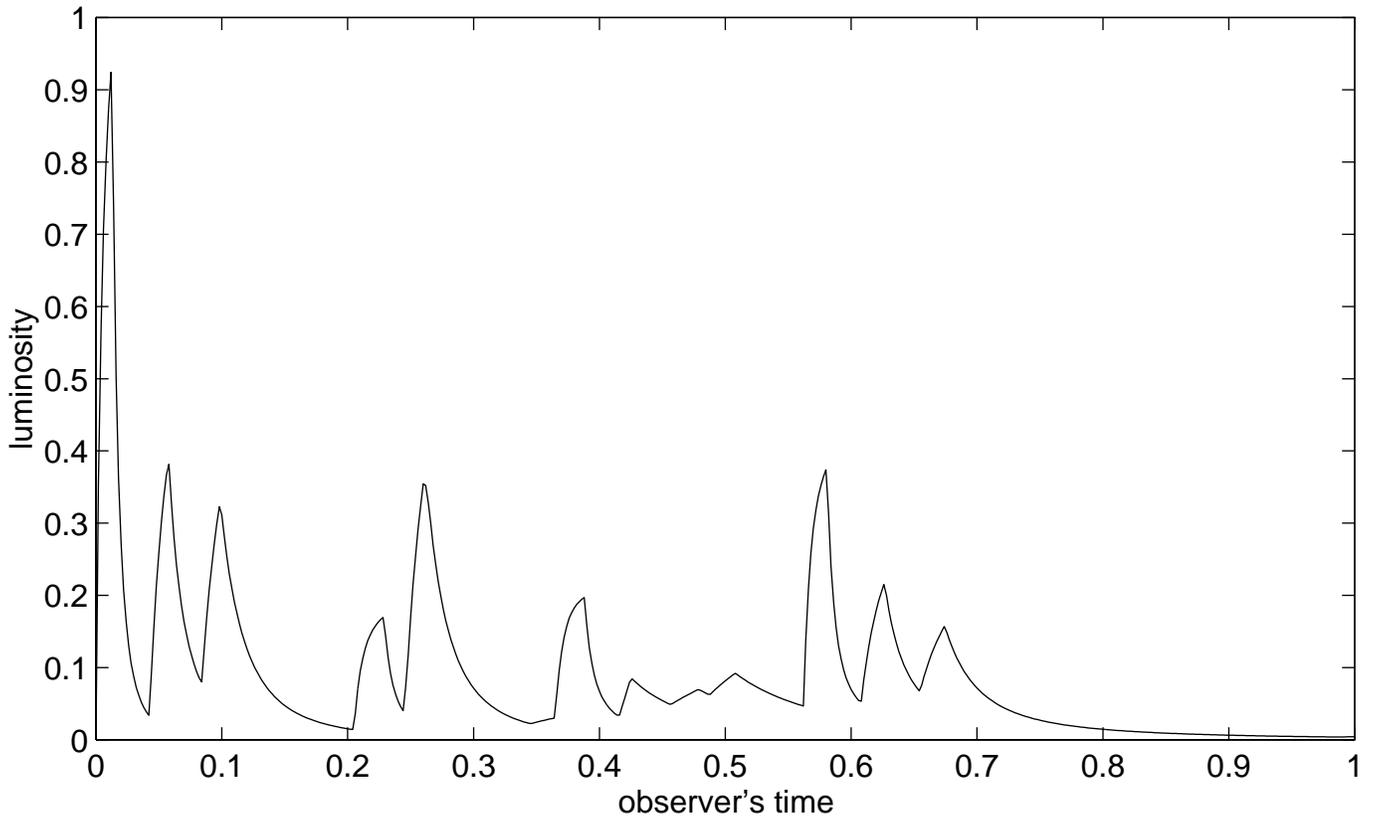
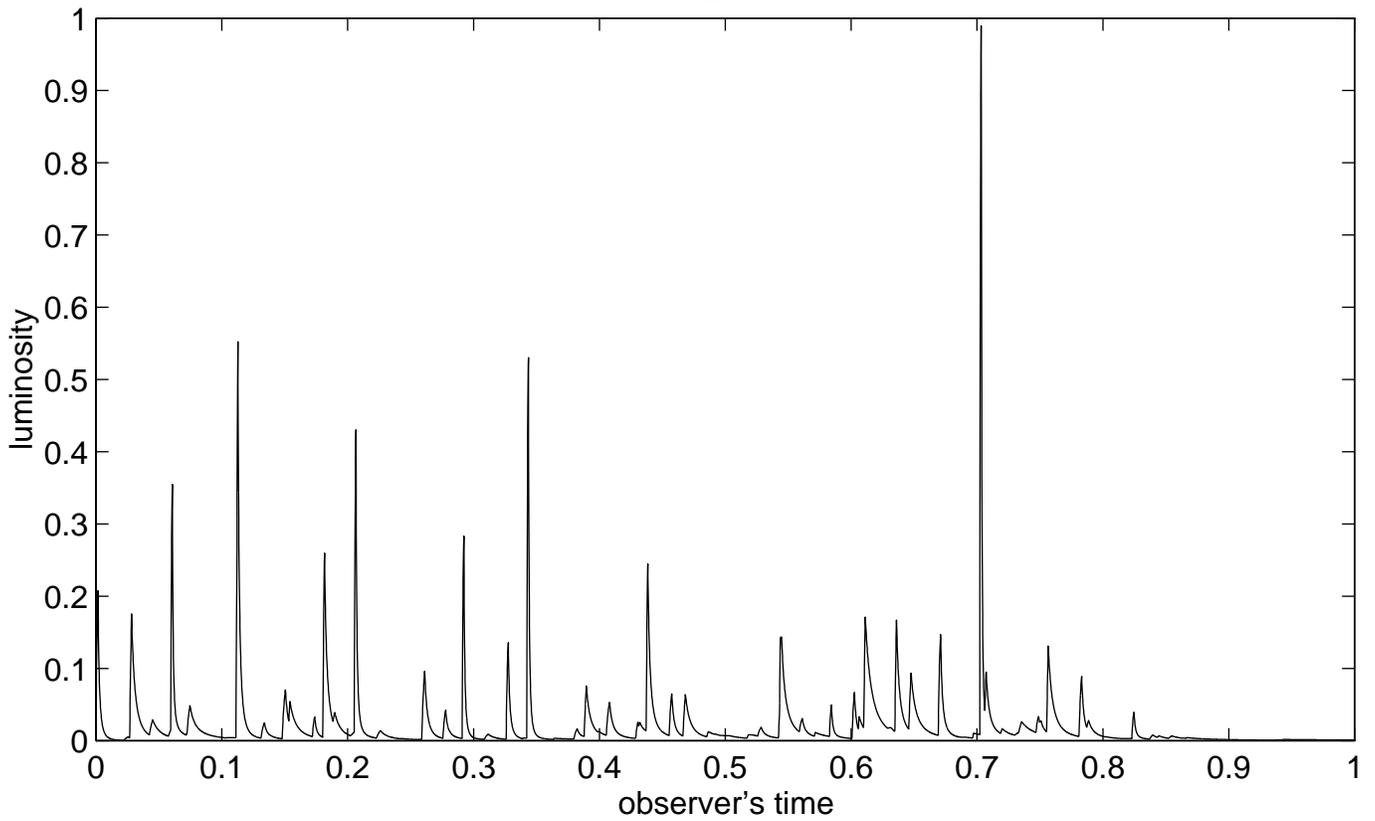
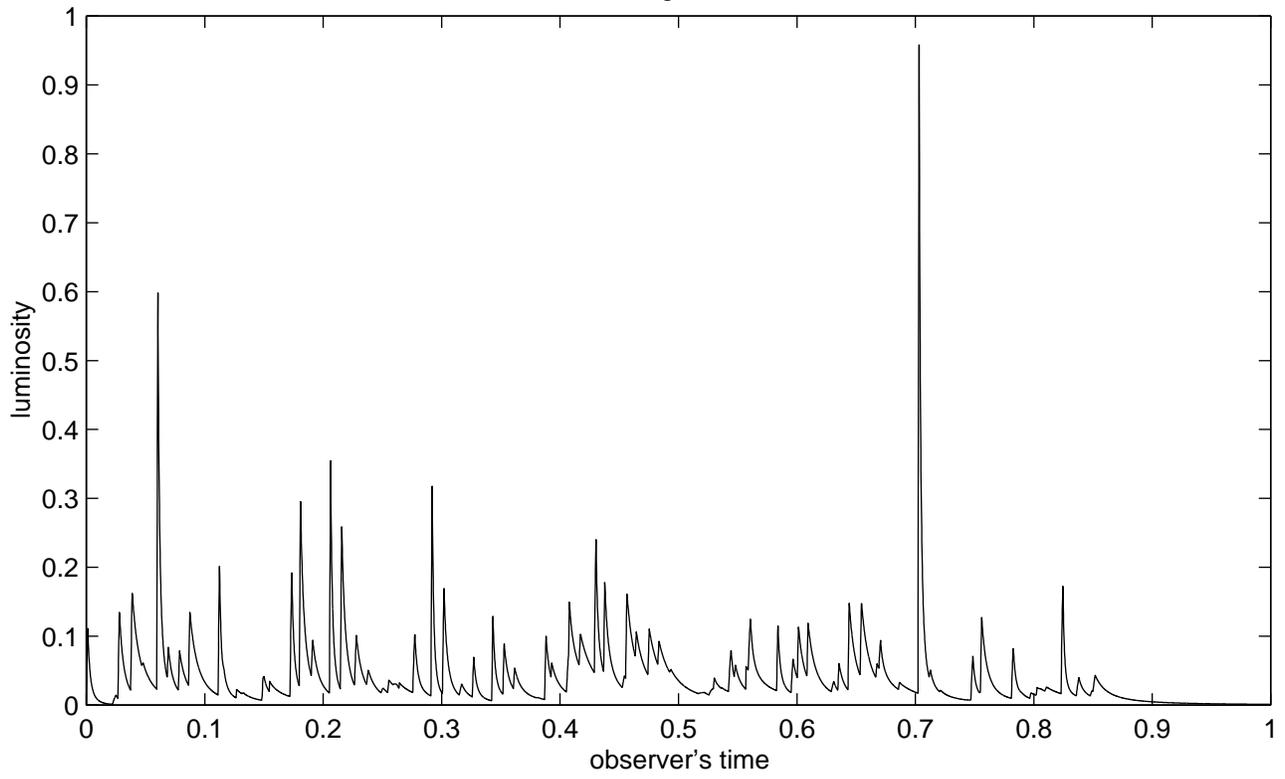
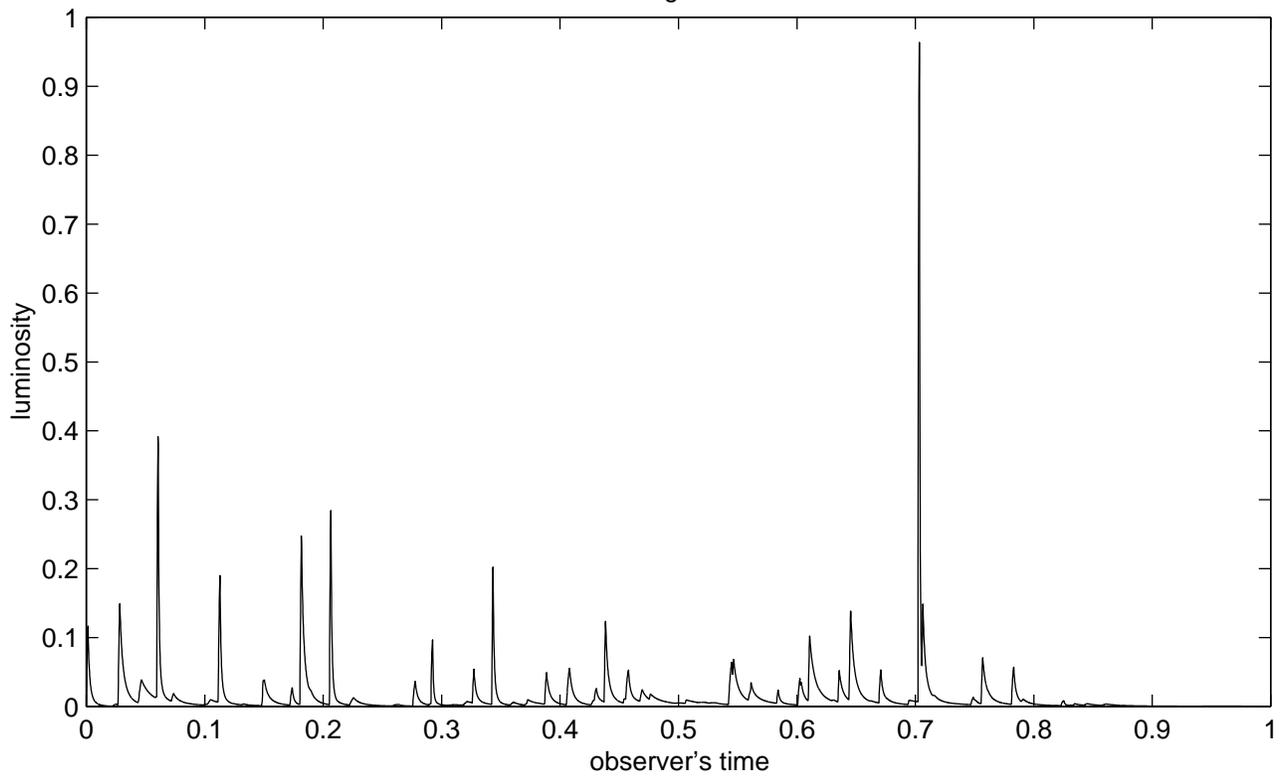
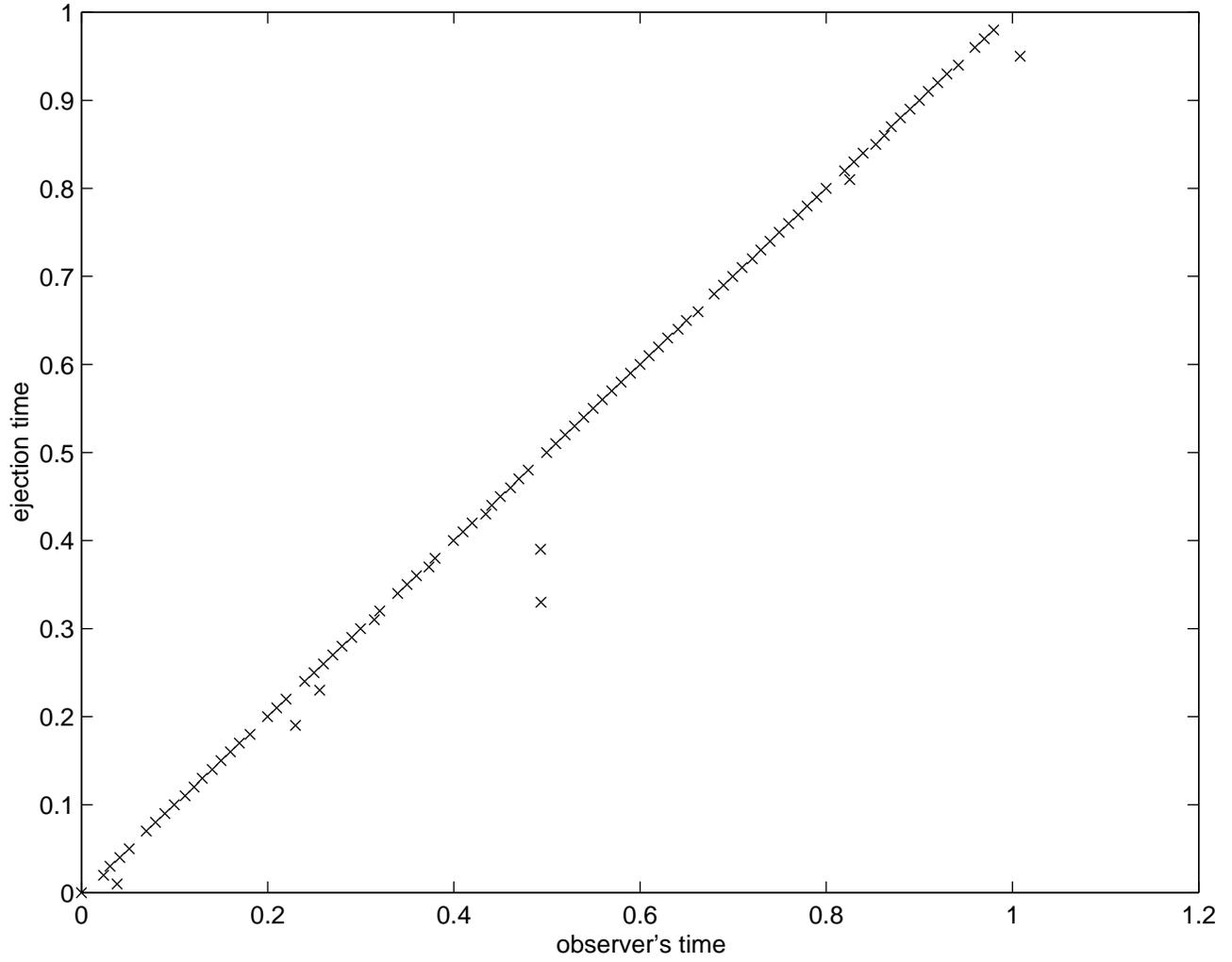
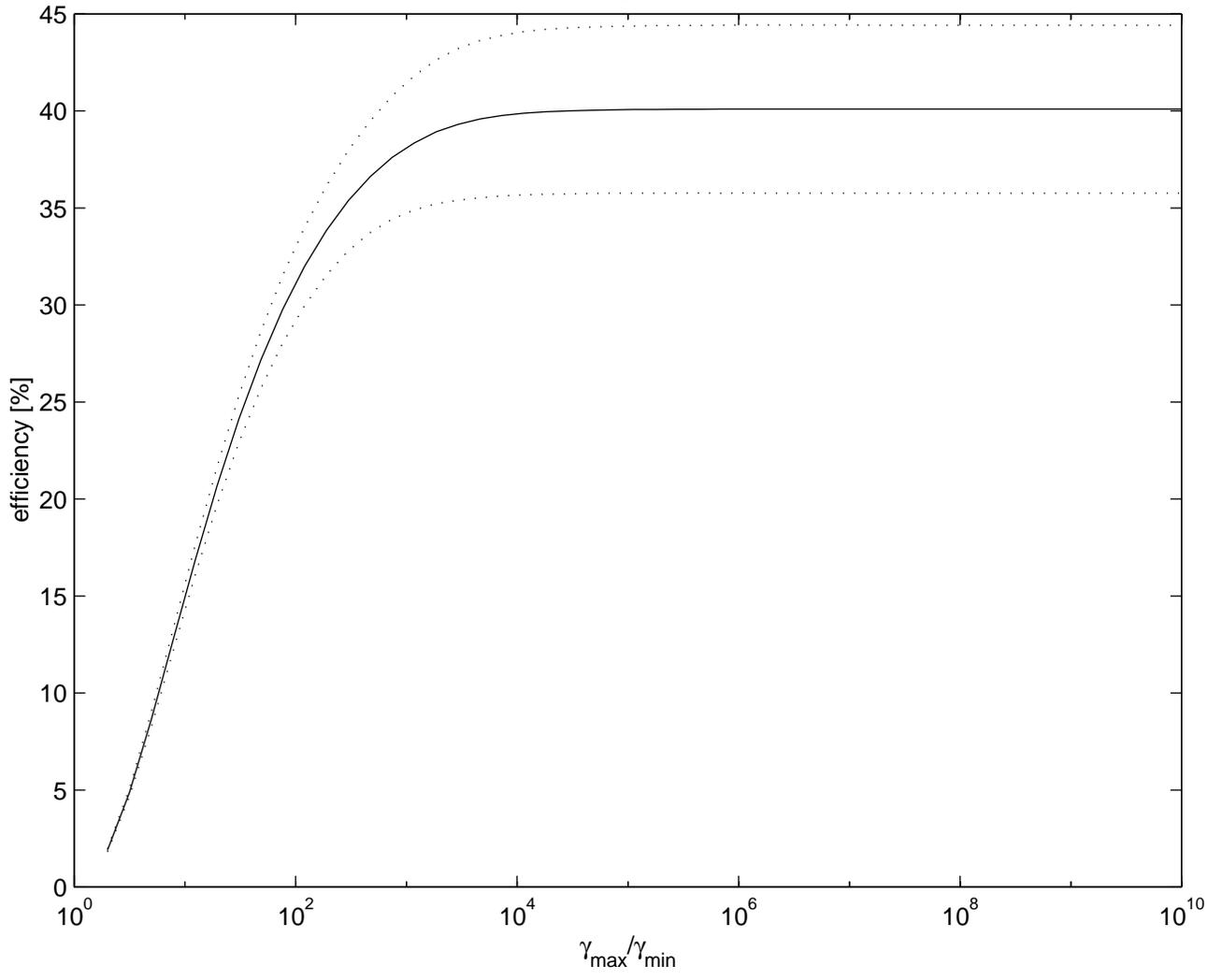
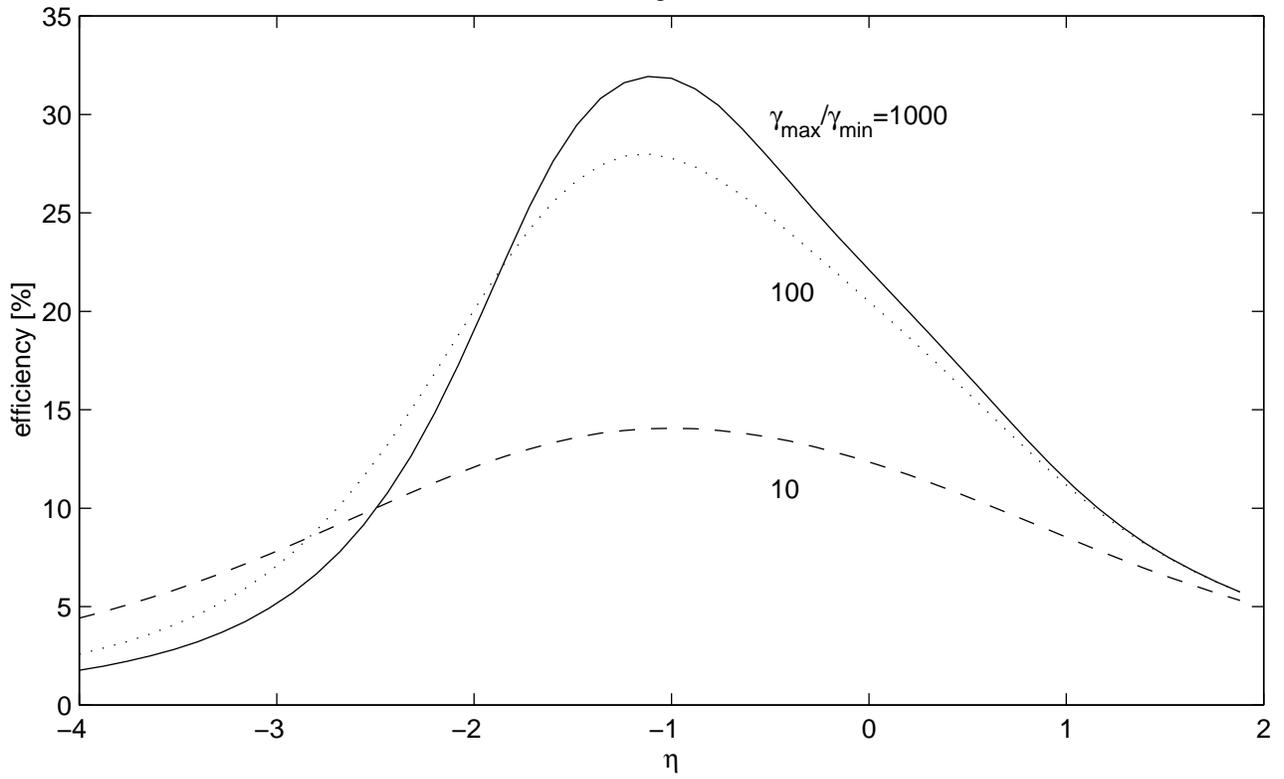
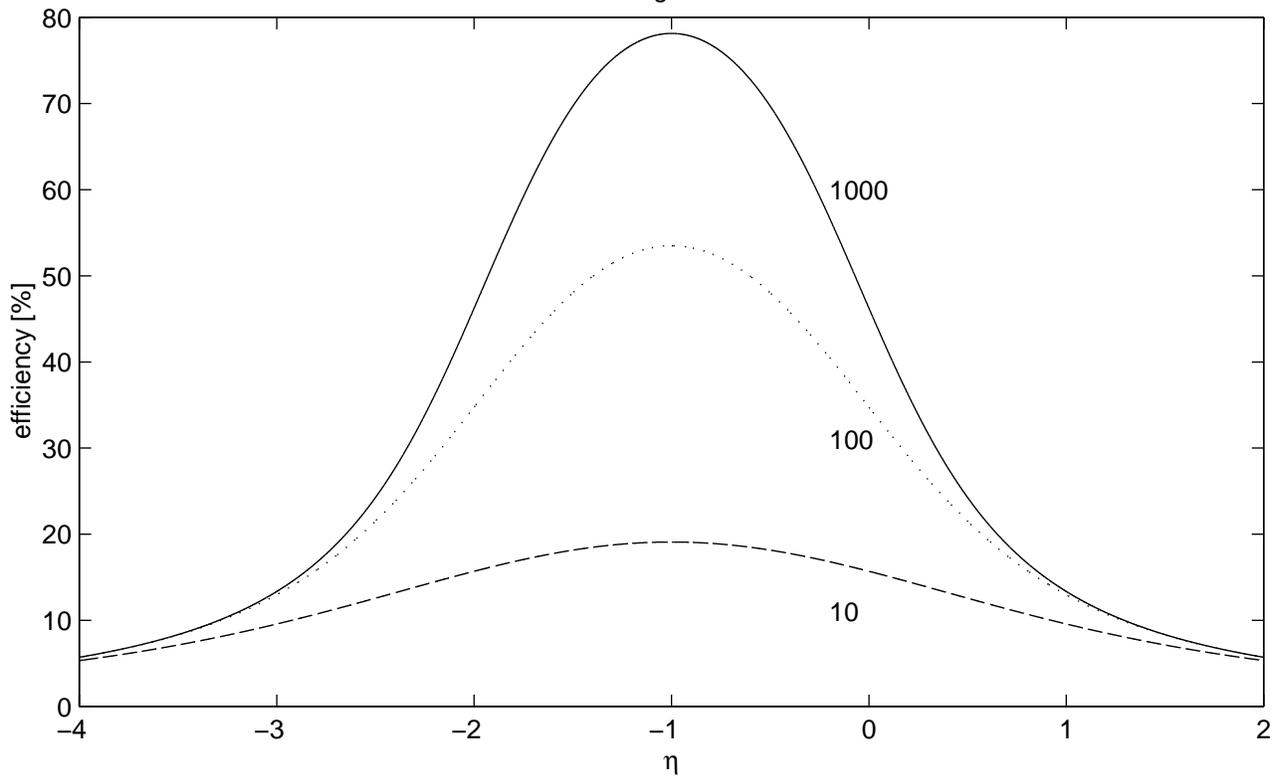
\end{document}